\documentclass[12pt]{article}
\newcommand{\bc}{\mbox{\boldmath$c$}}
\newcommand{\bl}{\mbox{\boldmath$l$}}
\newcommand{\bpi}{\mbox{\boldmath$\pi$}}
\newcommand{\bphi}{\mbox{\boldmath$\phi$}}
\newcommand{\bpsi}{\mbox{\boldmath$\psi$}}
\newcommand{\pbpi}{\,^{+}\!\mbox{\boldmath$\pi$}}
\newcommand{\mbpi}{\,^{-}\!\mbox{\boldmath$\pi$}}
\newcommand{\pmbpi}{\,^{\pm}\!\mbox{\boldmath$\pi$}}
\newcommand{\btau}{\mbox{\boldmath$\tau$}}
\newcommand{\pmbtau}{\,^{\pm}\!\mbox{\boldmath$\tau$}}
\begin{document}

\title{From areas to lengths in quantum Regge calculus}
\author{V.M.Khatsymovsky \\
 {\em Budker Institute of Nuclear Physics} \\ {\em
 Novosibirsk,
 630090,
 Russia}
\\ {\em E-mail address: khatsym@inp.nsk.su}}
\date{}
\maketitle
\begin{abstract}
Quantum area tensor Regge calculus is considered, some
properties are discussed. The path integral quantisation is
defined for the usual length-based Regge calculus
considered as a particular case (a kind of a state) of the
area tensor Regge calculus. Under natural physical
assumptions the quantisation of interest is practically
unique up to an additional one-parametric local factor of
the type of a power of $\det\|g_{\lambda\mu}\|$ in the
measure. In particular, this factor can be adjusted so that
in the continuum limit we would have any of the measures
usually discussed in the continuum quantum gravity, namely,
Misner, DeWitt or Leutwyler measure. It
is the latter two cases when the discrete measure turns out
to be well-defined at small lengths and lead to finite
expectation values of the lengths.
\end{abstract}
\newpage
The present paper combines results of the previous author's
works \cite{Kha,Kha1} to derive quantisation of the usual
length-based Regge calculus from quantisation of the area
Regge calculus model in the framework of the path integral
and canonical quantisation approaches. Namely, area Regge
calculus model
has been quantised and shown to lead to finite expectation
values of areas in \cite{Kha}. In the same Letter a
suggestion has been made and developed in \cite{Kha1} that
quantisation of Regge calculus is defined as that induced
from the more general Regge-like system with independent
4-simplex linklengths. In terms of metric, we issue from
the superspace of the metrics $\{g\}$ discontinuous on the
3-faces. The usual Regge calculus follows by imposing
continuity conditions describing a hypersurface
$\Gamma_{\rm cont}$ in this superspace. The functionals
$\psi (\{g\})$ on Regge calculus hypersurface $\Gamma_{\rm
cont}$ are naturally mapped into the set of the functionals
$\Psi (\{g\})$ on the original superspace of discontinuous
metrics,
\begin{equation}                                         
\Psi (\{g\})=\psi (\{g\})\delta_{\rm cont}(\{g\})
\end{equation}

\noindent where $\delta_{\rm cont}(\{g\})$ is a
$\delta$-function with support on $\Gamma_{\rm cont}$ so
that the quantum measure viewed as a functional on the
space of functionals is the result of the pull-back of such
the embedding,
\begin{equation}                                         
\mu_{\rm cont}(\cdot )
=\mu (\delta_{\rm cont}(\{g\})~\cdot ).
\end{equation}

\noindent In \cite{Kha1} it has been
shown that requirement that continuity conditions be
imposed
in a "face-independent" manner fixes $\delta_{\rm cont}$
uniquely. The term "face-independent" means that the
$\delta$-function factor responsible for the continuity of
the metric across a face depends only on the plane formed
by the face, not on the form and size of the face,

The situation can be represented by the following
(noncommutative) diagram,

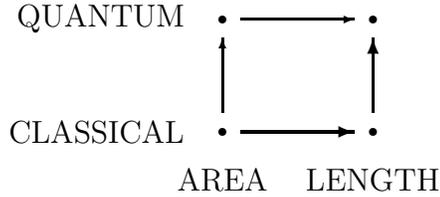
\vspace{-5mm}
\begin{picture}(80,40)(-20,-5)
\put(15,10){\circle*{1}}
\put(35,10){\circle*{1}}
\put(15,25){\circle*{1}}
\put(35,25){\circle*{1}}
\put(10,10){\makebox(0,0)[r]{CLASSICAL}}
\put(10,25){\makebox(0,0)[r]{QUANTUM}}
\put(15,5){\makebox(0,0)[t]{AREA}}
\put(35,5){\makebox(0,0)[t]{LENGTH}}
\thinlines
\put(15,12.5){\vector(0,1){10}}
\put(17.5,25){\vector(1,0){15}}
\thicklines
\put(35,12.5){\vector(0,1){10}}
\put(17.5,10){\vector(1,0){15}}
\end{picture}
\vspace{-5mm}

\noindent where lower path shown by thick arrows means
conventional approach to quantisation. Our approach is
shown by thin upper line. Using some analogy, we can say
that Regge calculus is considered as a kind of the state of
the area tensor Regge calculus.

The result of \cite{Kha} takes the form
\begin{eqnarray}
\label{VEV}
<\Psi (\{\pi\},\{\Omega\})> & = & \int{\Psi (-i\{\pi\},
\{\Omega\})\exp{\left (-\!
\sum_{\stackrel{t-{\rm like}}{(ABC)}}{\tau
_{(ABC)}\circ
R_{(ABC)}(\Omega)}\right )}}\nonumber\\
 & & \hspace{-20mm} \exp{\left (i
\!\sum_{\stackrel{\stackrel{\rm not}{t-{\rm like}}}{(ABC)}}
{\pi_{(ABC)}\circ
R_{(ABC)}(\Omega)}\right )}\prod_{\stackrel{\stackrel{\rm
 not}{t-{\rm like}}}{(ABC)}}d^6
\pi_{(ABC)}\prod_{(ABCD)}{{\cal D}\Omega_{(ABCD)}}
\nonumber\\
& \equiv & \int{\Psi (-i\{\pi\},\{\Omega\}){\rm d}
\mu_{\rm area}(-i\{\pi\},\{\Omega\})}.                   
\end{eqnarray}

\noindent Here $A\circ B$ $\stackrel{\rm def}{=}$ ${1\over
2}A^{ab}B_{ab}$. The field variables are area tensors
$\pi_{(ABC)}$ on the triangles $(ABC)$ (the 2-simplices
$\sigma^2$) and SO(4) connection matrices $\Omega_{(ABCD)}$
on the
tetrahedrons $(ABCD)$ (the 3-simplices $\sigma^3$). This
looks as the usual field-theoretical path
integral expression with exception of the following three
points. First, occurence of the Haar measure on the group
SO(4) of connections on separate 3-faces ${\cal D}\Omega
_{(ABCD)}$ connected with the specific form of the kinetic
term $\pi_{(ikl)}\circ{\Omega}^{\dag}_{(ikl)}\dot{\Omega}
_{(ikl)}$ if one passes to the continuous limit along any
of the coordinate direction chosen as time. Such form of
the kinetic term in the general 3D discrete gravity model
has been deduced by Waelbroeck \cite{Wae}.

Second, the sum of the terms $\pi_{(ABC)}\circ R_{(ABC)}$
in the exponential whereas the
exact connection representation of the Regge action is the
sum of the terms with 'arcsin' of the type $|\pi |\arcsin{(
\pi\circ R/|\pi |)}$ \cite{Kha2}. The matter is in the way
the eq. (\ref{VEV}) has been obtained. Namely, we issue
from the requirement that Feynman path integral based on
the canonical quantisation prescription should arise
whenever we pass to the continuous time limit along
whichever coordinate taken as time. The phase factor in the
integral is defined by the constraints. The latter follow
from the equations of motion which lead in the case of area
Regge calculus to the trivial solution $R$ $-$ $\bar{R}$ =
0, and the constraints are the same as if 'arcsin' were
omitted (in empty space; situation might be much more
complicated in the presence of matter fields!). The
exponential in (\ref{VEV}) just originates from the sum of
the constraints times Lagrange multipliers and is not the
original action.

Finally, third, occurence of a set of the triangles over
area tensors $\tau_{(ABC)}$ of which integrations are
absent. It is the
set of those triangles the curvature matrices on which are
functions, via Bianchi identities, of all the rest
curvatures. (Integrations over them might result in the
singularities of the type of $[\delta(R-\bar{R})]^2$). The
eq. (\ref{VEV}) displays the simplest choice of this set.
Namely, choose any of the coordinates $t$ and take all the
$t$-like triangles. The notion "$t$-like" suggests a
certain regular structure \cite{MisThoWhe} of the Regge
manifold considered as consisting of sequence of the 3D
Regge manifolds $t$ = $const$ of the same structure usually
called {\it the leaves of the foliation} along $t$. That
$t$-like triangles turn out to be just those ones on which
the curvature matrices can be expressed in terms of those
matrices on the other, contained in the 3D leaves
("leaf" triangles) and diagonal triangles might seem to be
accidental, but it seems quite interesting to connect
1-dimensionality of the coordinates along which quantum
fluctuations are absent ("time") with Bianchi identities.

The cutoff and positivity properties of the measure
(\ref{VEV}) can be seen for the trivial choice of the
$t$-like area tensors $\tau$ = 0. In this case the measure
in the eq. (\ref{VEV}) splits into the product of the
measures over separate (not $t$-like) triangles of the type
\begin{equation}                                         
\label{factor}
\label{separate}
\exp{(i\pi\circ R)}d^6\pi{\cal D}R.
\end{equation}

\noindent In turn, this splits into self- and
antiselfdual counterparts each taking the form
\begin{equation}                                         
\label{3d-eucl}
\exp{\left(i\bl\bphi{\sin{\phi}\over\phi}\right)}
d^3\bl{\sin^2(\phi /2) \over 4\pi^2\phi^2}d^3\bphi
\end{equation}

\noindent where $\bl$, $\bphi$ are 3D vectors into which
selfdual (or antiselfdual) parts of $\pi$ and generator of
$R$, respectively, are mapped. If applied to a (analitical)
function $f(-i\bl)$ this results in
\begin{equation}                                         
\label{exp-eucl}
<f(\bl)>=\int{f(\bl)\nu(l){d^3\bl\over 4\pi l^2}},~~~~
\nu (l)={2l \over \pi}\int_{0}^{\pi}{\exp{\left( -{l
\over\sin{\varphi}} \right)}d\varphi}.
\end{equation}

\noindent Here the two remarks are in order. First, one
could proceed directly from the Lorentzian signature case.
(This concerns the local frame indices, and we shall
distinguish between the terms "timelike", "spacelike" and
"$t$-like", "leaf", "diagonal" referred to the local frame
index and to the Regge analog of the world index,
respectively.) Considering effect of integrations over the
Lorentz generators $\bpsi$ = $(\psi^{01},\psi^{02},
\psi^{03})$ of rotations around the spacelike area
components with vector $\bl$ we find analogous to
(\ref{3d-eucl}) expression                               
\begin{equation}
\label{3d-loren}
\exp{\left(i\bl\bpsi{\sinh{\psi}\over\psi}\right)}
d^3\bl{\sinh^2(\psi /2) \over 4\pi^2\psi^2}d^3\bpsi
\end{equation}

\noindent with hyperbolic functions. Note that area spanned
by a triangle is described by the dual to $\pi$, and the
term "timelike" or "spacelike" area components just refers
to this object. Now (\ref{3d-loren}) should be
applied directly to $f(\bl)$ with real $\bl$. The result is
the same, eq. (\ref{exp-eucl}). The difficulty arises when
we incorporate the rest of the rotations, the
Euclidean ones, and get {\it complex} selfdual and
antiselfdual parts when trying to factorise anyhow.
Therefore we prefer to pass to integration over
imaginary $\psi^{0i}$ = $i\phi^{0i}$, i. e. over completely
Euclidean rotations. At the same time, we pass to the
purely imaginary variables characterising spacelike area
components ($\bl$ in (\ref{3d-loren})) leaving timelike
components unchanged. This just corresponds to making Wick
rotation of the timelike vector components together with
the overall rescaling by $-i$ area tensors over which
integration is made, as mentioned above. Note that we make
these redefinitions only over the field variables which
are, in turn, dummy variables in the path integral.
Necessity to make these over spacetime coordinates is
absent since Regge calculus is coordinateless formulation
($t$ simply numerates the leaves). May be, this solves the
problem of Wick rotation in quantum gravity?

Second remark concerns taking into account contribution of
the nonzero $t$-like triangles $\tau_{(ABC)}$. This is
necessary in order to describe evolution in $t$. Meanwhile,
corresponding terms in the exponential violate
factorisation of the measure (\ref{VEV}) into the
"elementary" ones (\ref{factor}) making calculation in
closed form hardly achievable. Since integrals considered
are only conditionally convergent, the properties of
analyticity at $\tau$ = 0 and of positivity are not quite
evident. To get some idea of possible form of the answer
take some simplified model of dependence on $\bphi$ of the
additional terms in the exponential of the measure. Whereas
$\sin{\phi}$ characterises antisymmetric part of the
curvature matrix, it's symmetric part containes
$\cos{\phi}$ and just arises in the considered terms due to
the dependence on the given curvature via Bianchi
identities. Therefore perform the following calculation,
\begin{eqnarray}
\int{f(-i\bl)\exp{\left(i\bl\bphi{\sin{\phi}\over\phi}
-\varepsilon\cos{\phi}\right)}
d^3\bl{\sin^2(\phi /2)\over 4\pi^2\phi^2}d^3\bphi}
&=&\int{f(\bl)\nu_{\varepsilon}(l){d^3\bl\over 4\pi l^2}},
\nonumber\\
\nu_{\varepsilon}(l)={2l \over \pi}\int_{0}^{\pi}{\exp{
\left(-{l\over\sin{\varphi}}\right)}\cos{(\varepsilon\cot{
\varphi})}d\varphi}.&&
\end{eqnarray}                                           

\noindent The $\nu_{\varepsilon}(l)$ is positive. It admits
expansion in powers of $\varepsilon$, although each
subsequent term of this expansion possesses more singular
behaviour at small $l$. The same properties can be
conjectured for the exact measure and for it's dependence
on $\tau$.

Now proceed to reducing the described measure to the
usual length-based Regge calculus. Roughly speaking, area
tensor Regge calculus can be considered as a simplicial
theory with metric defined in each the 4-simplex
independently of others, i. e. discontinuous on the
3-faces. To be more precise, we have even more general
theory in which even the 4-simplex metric is defined
ambiguously. This is because to define metric in a given
4-simplex it is sufficient to know it's ten areas of the
2-faces; meanwhile, introducing tensors instead of scalar
areas we thus introduce some redundant scalars which can be
constructed of these tensors. Thus we need the constraints
which would enforce area tensors to be bivectors
constructed of the tetrad corresponding to a certain metric
inside a given 4-simplex. Denote corresponding
$\delta$-function factor in the measure
$\delta_{\rm tetrad}$. Besides that, the constraints are
required which "glue" together different 4-simplex metrics,
i. e. just the above $\delta$-function factor $\delta_{\rm
cont}$. The constraints on the scalar
areas have been discussed in \cite{Mak,MakWil}. In our case
of area tensors a new possibility arises to get a system of
bilinear constraints although at the price of extending the
set of variables \cite{Kha3}. In fact, the above notation
for area tensor $\pi_{(ABC)}$ refers to the frame of a
certain 4-simplex $(ABCDE)$ where this tensor is defined.
The more detailed notation we use is $\pi_{(ABC)DE}$. The
idea is to introduce for any triangle the set of area
tensors defined in all the 4-simplices $\sigma^4$
containing the given triangle $\sigma^2$. Correspondingly,
also integrations over new variables should be introduced
in the measure.

Let us pass to
the 3D leaf notations for the simplices respecting the
above mentioned division of the simplices into $t$-like and
other, spacelike and diagonal ones. The vertices of the 3D
leaf are $i$, $k$, $l$, \dots.
The $i^+$ means image in the next-in-$t$ leaf of the vertex
$i$ taken at the moment $t$. The
notation $(A_1A_2\ldots A_{n+1})$ means unordered
$n$-simplex with
vertices $A_1$, $A_2$, \ldots, $A_{n+1}$ (triangle at $n$
= 2) while that without parentheses means ordered simplex.
The 4-simplices are arranged into the 4-prisms between the
pairs of the neighbouring 3D leaves. Let the 4-prism with
bases $(iklm)$ and $(i^+k^+l^+m^+)$ consists of the
4-simplices $(ii^+klm)$, $(i^+kk^+lm)$, $(i^+k^+ll^+m)$,
$(i^+k^+l^+mm^+)$. Take, e. g., $(ii^+klm)$ and consider
area tensors of it's 6 2-faces containing vertex $i$. These
are $\pi_{(ikl)}$ ($\pi_{(ikl)i^+m}$ in more detailed
notations),
\begin{equation}                                         
\pi_{(ii^+k)}\stackrel{\rm def}{=}
\tau_{(ii^+k)}\stackrel{\rm def}{=}\tau_{ik}
\end{equation}

\noindent and the two
cycle permutations of $k$, $l$, $m$. The other 4 tensors
are expressible as algebraic sums of these ones ensuring
closure of the 3-faces, e. g.
\begin{equation}
\label{pi-sum}
\pi_{(i^+kl)}=\pi_{(ikl)}+\tau_{ik}-\tau_{il}.          
\end{equation}

\noindent The tetrad constraints take the
form
\begin{eqnarray}
\label{pi*pi}
&\pi_{(ikl)}*\pi_{(ikl)}=0,~~~\pi_{(ikl)}*\pi_{(ikm)}=0,&
\nonumber\\
&\pi_{(ikl)}*\tau_{ik}=0,~~~\pi_{(ikl)}*\tau_{il}=0,&
\nonumber\\
&\tau_{ik}*\tau_{ik}=0,~~~\tau_{ik}*\tau_{il}=0,&
\nonumber\\
&\dots {\rm 2~cycle~perm}(k,l,m)\dots,&\\               
\label{V-constr}
&\pi_{(ikl)}*\tau_{im}=\pi_{(ilm)}*\tau_{ik}=
\pi_{(imk)}*\tau_{il}(=V_{(ii^+klm)}),&                 
\end{eqnarray}

\noindent the total number of them being 20. Here $A*B$ =
${1\over 4}A^{ab}B^{cd}\epsilon_{abcd}$ for the two tensors
$A$, $B$. The two constraints (\ref{V-constr}) provide
unambiguous definition of the 4-volume $V_{(ii^+klm)}$.

Important is that now integrations over $\tau_{(ABC)}$
should also be included. This is because fixing these
tensors as parameters in the length-based framework put
severe restriction on area tensors $\pi$ of the leaf and
diagonal triangles. On the other hand, the reason that
these integrations were omitted is absent now. Namely,
integrals over $\pi$ cannot lead to $\delta$-functions of
curvatures because of the nontrivial preexponent factor. It
quite may be that all integrations over $\tau_{(ABC)}$ will
prove to be finite. However, for our purposes it is
important to keep $\tau_{(ABC)}$ small so that we could use
the possibility of factorisation into elementary measures
(\ref{factor}). But to do this it is sufficient to equate
to small values the appropriately chosen 4 scalars per
vertex composed of $\tau$ components. For example, let
these constraints at the vertex $i$ be defined in the
considered 4-simplex $(ii^+klm)$. Thereby restrictions are
imposed on the linklength of $(ii^+)$, discrete analog of
the lapse-shift vector. The product over all the vertices
$\sigma^0$, $\delta_{\rm lapse-shift}$ serves to restrict
all the $t$-like links.
\begin{eqnarray}
\label{delta-i}                                         
\delta_{\rm lapse-shift}(i)=
\delta(\tau_{ik}\circ\tau_{ik}-\varepsilon^2_1)
\delta(\tau_{il}\circ\tau_{il}-\varepsilon^2_2)
\delta(\tau_{im}\circ\tau_{im}-\varepsilon^2_3)
\delta(\tau_{ik}\circ\tau_{il}-\zeta\varepsilon_1
\varepsilon_2),&&\\
\delta_{\rm lapse-shift}=\prod_{\sigma^0}               
{\delta_{\rm lapse-shift}(\sigma^0)}.\hspace{85mm}&&
\end{eqnarray}

\noindent Proceeding from the answer (finite expectation
values for the leaf and diagonal linklengths) the
(\ref{delta-i}) means smallness of the $(ii^+)$ linklength
and thereby smallness $O(\varepsilon)$ of all tensors of
the type $\tau_{i\dots}$. This corresponds to fixing
lapse-shift vector in the continuum general relativity.

Evidently, there is one-to-one correspondence of the
constraints (\ref{pi*pi}-\ref{V-constr}) to those on area
tensors $\pi^{ab}_{\lambda\mu}$ in the continuum theory at
a given point. It is possible by temporarily (in the given
4-simplex) redenoting $\pi_{(ikl)}$ $\to$ $\pi_{12}$,
$\tau_{ik}$ $\to$ $\pi_{01}$, \dots ${\rm 2~cycle~perm}
(i,k,l)$ and $(1,2,3)$\dots to rewrite (\ref{pi*pi}
-\ref{V-constr}) in the form
\begin{equation}                                        
\label{tetrad}
\epsilon_{abcd}\pi^{ab}_{\lambda\mu}\pi^{cd}_{\nu\rho}
\sim\epsilon_{\lambda\mu\nu\rho}.
\end{equation}

\noindent The constraints are in the form covariant w.r.t.
the world index. This is essential if we require that the
theory would have continuum limit measure invariant w. r.
t. the world index. However, usually this requirement is
relaxed, and local continuum measure is allowed to be not
necessarily invariant w. r. t. the conformal degree of
freedom. Therefore a power of
$\det{g}$, $g$ being metric tensor, or $\det{e}$, $e$ being
tetrad, is accepted as additional factor. This gives the
general form of the $\delta$-factor taking the constraints
into account. The product of such factors over all the
4-simplices just gives $\delta_{\rm tetrad}$.
\begin{equation}                                        
\label{delta-tetrad}
\delta_{\rm tetrad}(ii^+klm)=\int{V^{\eta}\delta^{21}
(\pi_{\lambda\mu}*\pi_{\nu\rho}-V\epsilon_{\lambda\mu
\nu\rho})dV},~~~\delta_{\rm tetrad}=\prod_{\sigma^4}{
\delta_{\rm tetrad}(\sigma^4)},
\end{equation}

\noindent $\eta$ being a parameter.

To study convergence properties of the integrals it is
convenient to make selfdual-antiselfdual splitting, in
particular, $\pi_{\lambda\mu}$ maps into 3-vectors $\pbpi
_{\lambda\mu}$, $\mbpi_{\lambda\mu}$. Here convenient are
3D notations $\epsilon_{\alpha\beta\gamma}\pmbpi^{\gamma}$
$\stackrel{\rm def}{=}$ $\pmbpi_{\alpha\beta}$, $\pmbtau
_{\alpha}$ $\stackrel{\rm def}{=}$ $\pmbpi_{0\alpha}$. Then
the constraints (\ref{tetrad}) yield
\begin{equation}                                        
\pmbtau_{\alpha}=\epsilon_{\alpha\beta\gamma}c^{\beta}
\pmbpi^{\gamma}+{1\over 2}C\epsilon_{\alpha\beta\gamma}
\pmbpi^{\beta}\times\pmbpi^{\gamma}.
\end{equation}

\noindent The selfdual
and antiselfdual area vectors in a given 4-simplex differ
by an overall rotation. Therefore choose as independent
variables $\bpi^{\alpha}$ (under which we mean $\pbpi
^{\alpha}$ for definiteness), $c^{\alpha}$, $C$ and SO(3)
rotation ${\cal O}$ which connects antiselfdual and
selfdual sectors, the overall number 16 just corresponds to
the
number of the tetrad components. Integrating over all the
36 4-simplex components $\pi^{ab}_{\lambda\mu}$ we find
\begin{equation}                                        
\label{int-tetrad}
\int{(\cdot)\delta_{\rm tetrad}(ii^+klm){\rm d}^{36}\pi}
=\int{(\cdot)C^{\eta-6}[\bpi^1\times\bpi^2\cdot\bpi^3]
^{\eta-6}{\rm d}C{\rm d}^3\bc{\rm d}^9\bpi{\cal DO}}.
\end{equation}

\noindent Consider for a moment continuum limit. It has
been proven \cite{Kha4} that modulo partial use of the eqs.
of motion our area tensor Regge calculus model results in
the continuum limit in the area-generalised
Hilbert-Palatini form of GR such that upon postulating the
tetrad form of area tensors we get usual GR. Therefore
integration over ${\cal D}\Omega$ by stationary phase
method just results in the integration over infinitesimal
connections ${\rm d}^{24}\omega^{ab}_{\lambda}$ in the
functional integral with Hilbert-Palatini form of the
action yielding an additional factor $V^{-6}$ = $C^{-6}
[\bpi^1\times\bpi^2\cdot\bpi^3]^{-6}$ = $(\det{\|g_{\lambda
\mu}\|})^{-3}$ in (\ref{int-tetrad}). On the functionals of
purely metric $g_{\lambda\mu}$ this leads to
the measure $(\det{\|g_{\lambda\mu}\|})^{{\eta -13\over 2}}
{\rm d}^{10}g_{\lambda\mu}$. It is Misner measure
\cite{Mis} at $\eta$ = 8 or the DeWitt measure \cite{DeW}
at $\eta$ = 13. It is also easy to define $\delta_{\rm
tetrad}$ so that in the continuum limit we would have the
Leutwyler measure
$(\det{\|g_{\lambda\mu}\|})^{-{3\over
2}}g^{00}{\rm d}^{10}g_{\lambda\mu}$ \cite{Leu,FraVil}.
Namely, we
should insert $[\bpi^1\times\bpi^2\cdot\bpi^3]$ into
(\ref{int-tetrad}) at $\eta$ = 8 for that.

Now in the discrete framework integrating over connections
we get for each leaf or diagonal triangle $\sigma^2$ the
tensor of which in (\ref{VEV}) is defined in the given
4-simplex the additional factor in (\ref{int-tetrad}),
\begin{equation}                                        
\label{cut-factor}
{\nu (|\pbpi_{\sigma^2}|)\over |\pbpi_{\sigma^2}|^2}
{\nu (|\mbpi_{\sigma^2}|)\over |\mbpi_{\sigma^2}|^2}
={\nu (|\bpi_{\sigma^2}|)^2\over |\bpi_{\sigma^2}|^4}.
\end{equation}

\noindent Possibility to get the factor in this form is
connected with factorisation into simple measures
(\ref{factor}) possible due to $\varepsilon$ $\ll$ 1 in
$\delta_{\rm lapse-shift}$.

The expression (\ref{cut-factor}) varies at small
$|\bpi_{\sigma^2}|$ as $|\bpi_{\sigma^2}|^{-2}$. The
behaviour at small areas is essential for the area
expectation values be vanishing or not. To see what
triangles to which 4-simplices can be attributed (i. e.
their tensors defined) in a regular way let us begin with
the leaf triangles and distribute these among the
3-simplices. Since $N^{(3)}_2$ = $2N^{(3)}_3$ where
$N^{(d)}_k$ is the number of $k$-simplices in the
$d$-dimensional Regge manifold we have two triangles in
each tetrahedron, e. g. let $(ikl)$ and $(ikm)$ be assigned
to $(iklm)$. Then define these in the frame of that one of
the two 4-simplices sharing $(iklm)$ which is in the future
in $t$, i. e. here in $(ii^+klm)$. The diagonal triangles
differing
by occurence of superscript $+$ on some of the vertices
will be assigned to the future 4-simplices in the same
4-prism. For example, $(i^+kl)$, $(i^+km)$ are defined in
$(i^+kk^+lm)$; $(i^+k^+l)$, $(i^+k^+m)$ are defined in
$(i^+k^+ll^+m)$; no triangles are defined in
$(i^+k^+l^+mm^+)$. Considered are the 4 4-simplices which
constitute a 4-prism. There are 6 triangles defined in the
4-simplices of any 4-prism (the triangle $(i^+k^+l^+)$
refers to the next-in-$t$ 4-prism). Because of
smallness of the lateral ($t$-like) 2-surface the areas of
the future in $t$ triangles are almost the same (differ by
$O(\varepsilon)$). Therefore the cutoff factor is
\begin{equation}                                        
\label{4-prism}
\left ({\nu (|\bpi_{\sigma^2_1}|)^2\over
|\bpi_{\sigma^2_1}|^4}\right )^3
\left ({\nu (|\bpi_{\sigma^2_2}|)^2\over
|\bpi_{\sigma^2_2}|^4}\right )^3+O(\varepsilon)
\end{equation}

\noindent for the 4-prism where $\sigma^2_1$, $\sigma^2_2$
are the triangles defined in the (base of) 4-prism.

It remains to insert $\delta_{\rm cont}$ which imposes
continuity conditions on the common 3-faces between the
4-simplices.
The
result of \cite{Kha1} looks as
\begin{equation}                                        
\delta_{\rm cont}=\prod_{\sigma^3}{V^4_{\sigma^3}\delta^6
(\Delta_{\sigma^3}S_{\sigma^3})}\left (\prod_{\sigma^2}
{V^3_{\sigma^2}\delta^3(\Delta_{\sigma^2}S_{\sigma^2})}
\right )^{-1}\prod_{\sigma^1}{V^2_{\sigma^1}\delta
(\Delta_{\sigma^1}S_{\sigma^1})}
\end{equation}

\noindent where $S_{\sigma^k}$ is induced metric on the
$k$-face $\sigma^k$ in the form of the edge components
\cite{PirWil}, i. e. $k{k+1\over 2}$ edge lengths squared
of the $k$-face, $\Delta_{\sigma^k}$ is discontinuity
across the $k$-face $\sigma^k$, $V_{\sigma^k}$ is the face
$k$-volume. Occurence of the factors with $k$ = 1, 2 serves
to cancel effect of the cycles enclosing the triangles and
leading to the singularities of the type of
$\delta$-function squared. The $\delta$-function in the
denominator means that the same function is contained in
the numerator and is thereby cancelled.

Let us reexpress $\delta_{\rm cont}$ in terms of
area tensor variables. Essential for us is division of the
set of $k$-faces into the two groups, $t$-like and not
$t$-like (leaf and diagonal) ones. Take the above 4-simplex
$(ii^+klm)$ and use the above local selfdual notations in
it. The face-independent $\delta$-factor responsible for
the metric continuity across the leaf 3-face $(iklm)$ takes
the form
\begin{equation}                                        
\label{3-leaf-cont}
[\bpi^1\times\bpi^2\cdot\bpi^3]^4\delta^6(\Delta_{(iklm)}
(\bpi^{\alpha}\bpi^{\beta})).
\end{equation}

\noindent Such the factor for the $t$-like 3-face
$(ii^+kl)$ looks analogously up to evident redenoting area
vectors,
\begin{equation}                                        
[\btau_1\times\btau_2\cdot\bpi^3]^4\delta (\Delta
(\bpi^3)^2)\delta (\Delta (\bpi^3\btau_1))\delta (\Delta
(\bpi^3\btau_2))\delta (\Delta (\btau_1)^2)\delta (\Delta
(\btau_2)^2)\delta (\Delta (\btau_1\btau_2)).
\end{equation}

\noindent Here $\Delta$ $\equiv$ $\Delta_{(ii^+kl)}$. The
$\delta$-factor for the $t$-like 2-face $(ii^+k)$ takes the
form (definition of the $\Delta_{\sigma^k}$ at $k$ $<$ 3 is
ambigious, $\Delta_{(ii^+k)}$ = $\Delta_{(ii^+k\tilde{l})}$
for some vertex $\tilde{l}$, and we assume that $\tilde{l}$
is just $l$).
\begin{equation}                                        
|\btau_1|^3|[\btau_1\times\btau_2\cdot\bpi^3]|^3\delta
(\Delta (\btau_1\times\btau_2)^2)\delta (\Delta
((\btau_1\times\btau_2)(\btau_1\times\bpi^3)))\delta
(\Delta (\btau_1\times\bpi^3)^2).
\end{equation}

\noindent Here $\Delta$ $\equiv$ $\Delta_{(ii^+k)}$. Write
out also $\delta$-factor for the $t$-like 1-face (link)
$(ii^+)$,
\begin{equation}                                        
\label{1-cont}
(\btau_1\times\btau_2)^2\delta (\Delta_{(ii^+)}(\btau_1
\times\btau_2)^2).
\end{equation}

Consider effect of inserting $\delta_{\rm cont}$ on the
expectation values. Perform simple power-counting estimate.
 Remarkable feature of $\delta_{\rm
cont}$ is that it is invariant w.r.t. the overall rescaling
area tensors of the $t$-like and not $t$-like triangles
separately, see (\ref{3-leaf-cont}-\ref{1-cont}). This is
important since we have fixed the scale of the $t$-like
area tensors. Qualitatively, behaviour of the 4-simplex
measure upon taking into account $\delta_{\rm lapse-shift}$
and $\delta_{\rm tetrad}$ can be modelled by the
1-dimensional one on the interval $(0,\infty)$,
\begin{equation}                                        
\label{model}
x^{\eta -7}{\rm d}x,~~~e^{-x}x^{\eta -9}{\rm d}x,~~{\rm or}
~~e^{-2x}x^{\eta -11}{\rm d}x
\end{equation}

\noindent if 0, 1 or 2 area tensors in (\ref{VEV}) are
defined in the given 4-simplex, respectively. Here $x$
models the scale of areas. If we adjust $\delta_{\rm
tetrad}$ to reproduce the
Leutwyler measure in the continuum limit
(cf. discussion
after eq. (\ref{int-tetrad})) this would be equivalent to
putting $\eta$ = 11 in (\ref{model}). Inserting
$\delta_{\rm cont}$ can be modelled by inserting the
scale-invariant functions $x\delta (\Delta x)$ so that the
two measures $f_1(x_1){\rm d}x_1$, $f_2(x_2){\rm d}x_2$
intended to model the two measures in the 4-simplices
sharing a $t$-like face map into
\begin{equation}                                        
f_1(x_1){\rm d}x_1x_1\delta (x_1-x_2)f_1(x_2){\rm d}x_2
\Longrightarrow
xf_1(x)f_2(x){\rm d}x.
\end{equation}

\noindent In the case of the two neighbouring 4-simplices
one being in the future in $t$ w.r.t. another one things
are somewhat more complicated because now
$\delta_{\rm cont}$ compares area tensors of the "earlier"
triangles of the "later" 4-simplex and area tensors of the
{\it "later"} triangles of the "earlier" 4-simplex. The
latter tensors are expressible as algebraic sums of those
ones taken above as independent set in the given 4-simplex,
(\ref{pi-sum}).
Modulo $\delta_{\rm lapse-shift}$ inserted, this eq. means
that
$\pi_{(i^+kl)}$ is $\pi_{(ikl)}$ up to $O(\varepsilon)$.
Therefore the two measures $f_1(x_1){\rm d}x_1$, $f_2(x_2)
{\rm d}x_2$ modelling the measures in the considered
4-simplices can be considered to map into
\begin{equation}                                        
f_1(x_1){\rm d}x_1\delta (\varepsilon -|x'_1-x_1|)x'_1
\delta (x'_1-x_2)f_1(x_2){\rm d}x_2
\Longrightarrow
xf_1(x)f_2(x){\rm d}x
\end{equation}

\noindent up to $O(\varepsilon)$. Here $\delta (\varepsilon
-|x'_1-x_1|)$ models the $\delta_{\rm lapse-shift}$. Taking
into account the 4-prism factor (\ref{4-prism}) the model
measure for $n$ 4-prisms reads
\begin{equation}                                        
\int{\left (e^{-6x}x^{4\eta -36}\right )^n{{\rm d}x\over
x}}.
\end{equation}

\noindent This gives for $n$ $\gg$ $j$
\begin{equation}                                        
\label{x-vac}
<x^j>=\left ({2\over 3}\eta -6\right )^j
\end{equation}

\noindent (but the number of the
prisms along $t$ should be much less than
$\varepsilon^{-1}$ to ensure possibility of neglecting
tensors $\tau$ in the exponential of (\ref{VEV}) and
possibility to treat leaf and diagonal tensors $\pi$ as
practically unchanged in $t$). The eq. (\ref{x-vac})
corresponds formally to the $\delta$-function-like limiting
($n$ $\to$ $\infty$) measure, but this is due to, first,
neglecting tensors $\tau$ ($\varepsilon$ $\to$ 0), second,
simplicity of the suggested 1D model of estimating. The
main outcome is that the length expectation values are
finite and nonzero at $\eta$ $>$ 9. This is the case for
the above choices of $\delta_{\rm tetrad}$ leading in the
continuum limit to the Leutwyler or
DeWitt measures. At $\eta$ $\leq$ 9, in particular, in the
Misner measure case, the length expectation values are
zero, and this is connected with an overall divergence of
the measure at small length scale.

To summarize, we have considered the following expression
for quantum measure for the usual length-based Regge
calculus,
\begin{equation}                                        
{\rm d}\mu_{\rm length}=
\left (\delta_{\rm lapse-shift}
\prod_{\stackrel{{t-{\rm like}}}{(ABC)}}d^6\tau_{(ABC)}
\right )\delta_{\rm tetrad}\delta_{\rm cont}{\rm d}
\mu_{\rm area}.
\end{equation}

\noindent This practically one-parametric (depending on the
choice of $\delta_{\rm tetrad}$) equation results in
quantisation of the general relativity in the continuum
(long-wavelength) limit with some usual local measure at
the same time providing finite expectation values of the
linklengths.

\bigskip

The present work was supported in part by the Russian
Foundation for Basic Research through Grant No.
03-02-17612, through Grant No. 00-15-96811 for Leading
Scientific Schools and by the Ministry of Education Grant
No. E00-3.3-148.


\begin{thebibliography}{99}
\bibitem{Kha}
 V.M. Khatsymovsky, Area expectation values in quantum area
 Regge calculus, gr-qc/0212110, to appear in Phys.Lett. B
\bibitem{Kha1}
 V.M. Khatsymovsky, Regge calculus from discontinuous
 metrics, gr-qc/0304006, submitted to Phys.Lett. B
\bibitem{Wae}
 H. Waelbroeck, {\it Class.Quant.Grav.~}{\bf 7}~(1990)~751.
\bibitem{Kha2}
 V.M. Khatsymovsky, {\it Class.Quant.Grav.~}{\bf 6}~(1989)~
 L249.
\bibitem{MisThoWhe}
 C.W. Misner, K.S. Thorne, J.A. Wheeler, Gravitation,
 San Francisco, 1973.
\bibitem{Mak}
 J. M\"{a}kel\"{a}, {\it Class.Quant.Grav.~}{\bf 17}~(2000)
 ~4991, gr-qc/9801022.
\bibitem{MakWil}
 J. M\"{a}kel\"{a}, R.M. Williams, {\it Class.Quant.Grav.~}
 {\bf 18}~(2001)~L43, gr-qc/0011006.
\bibitem{Kha3}
 V.M. Khatsymovsky, {\it Gen.Rel.Grav.~}{\bf 27}~(1995)
 ~583, gr-qc/9310004.
\bibitem{Kha4}
 V.M. Khatsymovsky, {\it Phys.Lett.~}{\bf 547B}~(2002)~321,
 gr-qc/0206067.
\bibitem{Mis}
 C.W. Misner, {\it Rev.Mod.Phys.~}{\bf 29}~(1957)~497.
\bibitem{DeW}
 B.S. DeWitt, {\it J.Math.Phys.~}{\bf 3}~(1962)~1073.
\bibitem{Leu}
 H. Leutwyler, {\it Phys.Rev.~}{\bf B134}~(1964)~1155.
\bibitem{FraVil}
 E.S. Fradkin, G.A. Vilkovisky, {\it Phys.Rev.~}{\bf D8}~
 (1974)~4241.
\bibitem{PirWil}
 T. Piran, R.M. Williams, {\it Phys.Rev.~}{\bf D33}
 ~(1986)~1622.
\end{thebibliography}
\end{document}